\documentclass[10pt,a4paper]{article}
\usepackage[utf8]{inputenc}
\usepackage[english]{babel}
\usepackage{amsmath}
\usepackage{amsfonts}
\usepackage{amssymb}
\usepackage{amsthm}
\usepackage{color}
\usepackage[left=2cm,right=2cm,top=2cm,bottom=2cm]{geometry}

\numberwithin{equation}{section}

\newtheorem{lemma}{Lemma}
\numberwithin{lemma}{section} 

\newtheorem{thm}{Theorem}
\numberwithin{thm}{section} 

\theoremstyle{remark}
\newtheorem{rem}{Remark}
\numberwithin{rem}{section}

\newcommand{\R}{\mathbb{R}}
\newcommand{\Sp}{\mathbb{S}}
\newcommand{\eqdef}{\overset{def}{=}}	

\newcommand{\tW}{\widetilde{W}}

\title{An iterative inversion of weighted Radon transforms along hyperplanes}
\author{F.O. Goncharov\thanks{CMAP, Ecole Polytechnique, CNRS, Universit\'{e} Paris-Saclay, 91128, Palaiseau, France; email: fedor.goncharov.ol@gmail.com}}

\begin{document}

\maketitle
\begin{abstract}
We propose iterative inversion algorithms for weighted Radon transforms $R_W$ along hyperplanes in $\R^3$. More precisely, expanding the weight $W=W(x,\theta), x\in \R^3, \theta\in \Sp^2,$ into the series of spherical harmonics in $\theta$ and assuming that the zero order term $w_{0,0}(x)\neq 0, x\in \R^3,$ we reduce the inversion of $R_W$ to solving  a linear integral equation. 
In addition, under the assumption that the even part of $W$ in $\theta$ (i.e., $\frac{1}{2}(W(x,\theta) + W(x,-\theta))$) is close to $w_{0,0}$, the aforementioned linear integral equation can be solved by the method of successive approximations. Approximate inversions of $R_W$ are also given. 
Our results can be considered as an extension to 3D of two-dimensional results of Kunyansky (1992), Novikov (2014), Guillement, Novikov (2014). In our studies we are 
motivated, in particular, by problems of emission tomographies in 3D.
In addition, we generalize our results to the case of dimension $n>3$.
\end{abstract}

\section{Introduction}
\par We consider the weighted Radon transforms $R_W$ defined by the formula
\begin{align}\label{weighted_radon_general_def}
	&R_Wf(s,\theta) = \int\limits_{x\theta = s} W(x,\theta) f(x)\, dx, \, (s,\theta) \in \R\times \Sp^{2}, \, x\in \R^3,
\end{align}
where $W = W(x,\theta)$ is the weight, $f = f(x)$ is a test function. 
\par  In this work we assume that 
\begin{align}\label{weight_func_assymp}
	&W\in C(\R^3 \times \Sp^{2}) \cap L^{\infty}(\R^3 \times \Sp^{2}),\\ \label{w0_def}
	& w_{0,0}(x) \eqdef \dfrac{1}{4\pi}\int\limits_{\Sp^{2}} W(x,\theta) \, d\theta, \, w_{0,0}(x)\neq 0, \, x\in \R^3,\\ \label{test_func_assymp}
	&f\in L^{\infty}(\R^3), \, \mathrm{supp}\, f \subset D,
\end{align}
where $W$ and $f$ are complex-valued, $d\theta$ is an element of standard measure on $\Sp^{2}$,
 $D$ is an open bounded domain (which is fixed apriori). 
\par If $W\equiv 1$, then $R_W$ is reduced to the classical Radon transform along hyperplanes in $\R^3$ introduced in \cite{radon1917}; see also, e.g., \cite{natterer1986mathematics}, \cite{denisiuk2016}.
\par For known results on the aforementioned transforms $R_W$ with non-constant $W$ 
we refer to \cite{quinto1983invertibility}, \cite{beylkin1984inversion}, 
\cite{boman1987support}, \cite{goncharov2016analog}. In particular, in \cite{quinto1983invertibility} it was shown that $R_W$ is injective on $L_0^p(\R^3), \, p\geq 2$ ($L^p$ functions on $\R^3$ with compact support) if $W\in C^2$ and is real-valued, strictly positive and satisfies the strong symmetry assumption of rotation invariancy (see \cite{quinto1983invertibility} for details). On the other hand, in \cite{boman1987support} it was also proved that $R_W$ is  injective if $W$ is real-analytic and strictly positive.
\par Besides, in \cite{beylkin1984inversion} the inversion of $R_W$ is reduced to 
solving a Fredholm type linear integral equation in the case of infinitely smooth strictly positive $W$ with the symmetry $W(x,\theta) = W(x,-\theta)$. 
\par In turn, the work \cite{goncharov2016analog} extends to the case of weighted Radon transforms along hyperplanes in multidimensions the two-dimensional Chang approximate inversion formula (see \cite{chang1978method}) and the related two-dimensional result of \cite{novikov2011weighted}. In particular, \cite{goncharov2016analog} describes all $W$ for which such Chang-type formulas are simultaneously explicit and exact in multidimensions.

\par We recall that inversion methods for $R_W$ admit tomographical applications in the framework of the scheme described as follows (see \cite{goncharov2016analog}). 
 

\par It is well-known that in many tomographies measured data are modeled by weighted ray transforms $P_wf$ defined by the formula
\begin{align}\label{Pw_def}
	&P_wf(x,\alpha) = \int\limits_{\R} w(x + \alpha t, \alpha)f(x + \alpha t)\, dt, \, (x,\alpha)\in T\Sp^2,\\
	&T\Sp^2 = \{(x,\alpha) \in \R^3 \times \Sp^2 : x\alpha = 0\} , \nonumber
\end{align}
where $f$ is an object function defined on $\R^3$, $w$ is the weight function defined on $\R^3\times \Sp^2$, and $T\Sp^2$ can be considered as the set of all rays (oriented straight lines) in $\R^3$, see, e.g., \cite{chang1978method}, \cite{natterer1986mathematics}, \cite{kunyansky1992generalized}, \cite{guillement2014inversion}. 
\par In particular,  in the single photon emission computed tomography (SPECT) the weight $w$ is given by the formulas  (see, e.g., \cite{natterer1986mathematics}): 
\begin{align}
\label{xray_weight}
	w(x,\theta) = \exp 
	\big(-\int\limits_{0}^{\infty}a(x + t\theta)\,dt), \, (x,\theta)\in \R^3\times \Sp^2,
\end{align}
where $a = a(x)$ is the attenuation coefficient.
\par Moreover, in \cite{goncharov2016analog} (Section 3) it was shown that if $P_wf$ are given for all rays parallel to some fixed plane $\Sigma$ in $\R^3$ then $R_Wf$ with appropriate $W$ can be obtained by the explcit formulas from $P_wf$ and $w$ (in a similair way with the case $w\equiv 1, \, W\equiv 1$, see Chapter 2, formula (1.1) of  \cite{natterer1986mathematics} and also \cite{grangeat1991mathematical}, \cite{denisiuk2016}). Therefore, reconstruction of $f$ from data modeled by $P_wf$, defined by \eqref{Pw_def} and restricted to all rays parallel to $\Sigma$, can be reduced to reconstruction of $f$ from $R_Wf$, defined by \eqref{weighted_radon_general_def}.
In \cite{goncharov2016analog} it was aslo indicated that the reduction from $P_wf$ to $R_Wf$ with subsequent reconstruction of $f$ from $R_Wf$ and $W$ can drastically reduce the impact of the random noise in the initial data modeled as $P_wf$. 
\par In the present work we continue studies of \cite{goncharov2016analog}, on one hand, and of \cite{kunyansky1992generalized}, \cite{novikov2014weighted}, \cite{guillement2014inversion}, on the other hand. In particular, we extend the two-dimensional results of \cite{kunyansky1992generalized}, \cite{novikov2014weighted}, \cite{guillement2014inversion} to the case of weighted Radon transforms along hyperplanes in multidimensions.
In particular, under assumptions \eqref{weight_func_assymp}, \eqref{w0_def}, expanding $W = W(x,\theta)$ into the series of spherical harmonics in $\theta$ we reduce the reconstruction of $f$ to solving a linear integral equation 
(see Section~\ref{sect_main_results}). In particular, if the even part of $W$ in $\theta$ (i.e., $\tilde{W}(x,\theta) = \frac{1}{2}(W(x,\theta) + W(x,-\theta))$) is close to $w_{0,0}$,  then such linear integral equation can be solved by the method of successive approximations (see Subsections~\ref{subsect_precise_sol},~\ref{subsect_finite_weight} for details). 
 
\par Note that our linear integral equation is very different from the aforementioned linear integral equation of \cite{beylkin1984inversion} (in particular, in our conditions 
on $\tilde{W}$, ensuring the applicability of the method of successive approximations).
\par Note also that in  \cite{chang1978method}, \cite{kunyansky1992generalized},  \cite{novikov2014weighted}, \cite{guillement2014inversion} the two-dimensional 
prototype of our inversion approach was developed in view of its numerical efficiency
in problems of emission tomographies, including good stability to strong random noise
in the emission data.\\

In more details our results can be sketched as follows.
\par We use the following expansion for $W$:
\begin{align}\label{approximation_weight_intro}
	&W(x,\theta(\gamma,\phi)) = \sum\limits_{k = 0}^\infty 
	\sum\limits_{n=-k}^{k}w_{k,n}(x) Y^{n}_{k}(\gamma,\phi), \, x\in \R^3, \\
	\label{Ykn_def}
	&Y_{k}^{n}(\gamma,\phi) \eqdef p_{k}^{|n|}(\cos\gamma)e^{in\phi},
	\, k\in \mathbb{N}\cup \{0\}, \, n = \overline{-k,k},\\
	\vspace{0.4cm}
	\label{theta_gamma_phi}
	&\theta(\gamma,\phi) = (\sin\gamma \cos\phi, \, \sin\gamma\sin\phi,\,  \cos\gamma)\in \Sp^2,\, \gamma \in [0,\pi], \, \phi\in [0,2\pi],
\end{align}
where $p_k^{n}(x), \, x\in [-1,1]$, are the associated semi-normalized Legendre polynomials. Polynomials $p_k^n$ are well-known in literature (see e.g. \cite{stein2016introduction}) and are defined using the ordinary Legendre polynomials $p_k$ by the formulas:
\begin{align}\label{legendre_positive}
	&p_k^{n}(x) = (-1)^{n}
	\sqrt{\dfrac{(k-n)!}{(k+n)!}}
	(1-x^2)^{n/2}\dfrac{d^n}{dx^n}(p_k(x)), \, n,k\in \mathbb{N}\cup \{0\},\\ 
	\label{legendre_ordinary}
	&p_k(x) = \dfrac{1}{2^k k!}\dfrac{d^k}{dx^k}[(x^2-1)^k],\, x\in [-1,1],
\end{align}
see also \cite{stein2016introduction}, \cite{zhizhiashvili1979fourier} for other properties of the associated Legendre polynomials. In addition, coefficients $w_{k,n}$ in \eqref{approximation_weight_intro} are defined by the formulas:
\begin{align}\label{wkn_def}
	&w_{k,n}(x) = c(k,n)\int\limits_{0}^{2\pi}d\phi \, e^{-in\phi}\int\limits_{0}^{\pi}W(x,\theta(\gamma,\phi)) p^{|n|}_{k}(\cos \gamma) \sin\gamma \, d\gamma,\\ \nonumber
	&c(k,n) = \dfrac{(2k+1)}{4\pi}, \, k\in \mathbb{N}\cup \{0\}, \, n = 0, \pm 1, \cdots, \pm k.
\end{align}

Under assumption \eqref{weight_func_assymp}, for each fixed $x$, series \eqref{approximation_weight_intro} converges in $L^2(\Sp^{2})$; see, e.g., \cite{stein2016introduction} (Chapter 4), \cite{morimoto1998analytic} (Chapter 2), \cite{zhizhiashvili1979fourier}.

\par We consider also
\begin{align}\label{sigma_WDm_def}
	&\sigma_{\tW,D,m} = \dfrac{1}{2\pi\sqrt{2}}\sum\limits_{k = 1}^{m}
	\sum\limits_{n=-2k}^{2k} 
		\sup_{x\in D}\left|
			\dfrac{w_{2k,n}(x)}{{w_{0,0}}(x)}
		\right| \text{ for } m\in \mathbb{N},\\ \nonumber
	&\sigma_{\tW,D,m} = 0 \text{ for } m=0,\\ \label{sigma_WDinfty_def}
	&\sigma_{\tW,D,\infty} = \lim\limits_{m\rightarrow \infty}\sigma_{\tW,D,m},\\ 	
	\label{wN_def}
	&W_N(x,\theta(\gamma,\phi)) = \sum\limits_{k = 0}^N 
	\sum\limits_{n=-k}^{k}w_{k,n}(x) Y_k^n(\gamma,\phi), \\
	\label{wN_tilde_def}
	&\widetilde{W}_N(x,\theta(\gamma,\phi)) = \sum\limits_{k = 0}^{[N/2]} 
	\sum\limits_{n=-2k}^{2k}w_{2k,n}(x) Y_{2k}^{n}(\gamma,\phi), \\ \nonumber
	&x\in \R^3, \, \gamma \in [0,\pi], \, \phi\in [0,2\pi],
\end{align}
where coefficients $w_{k,n}$ are defined in \eqref{wkn_def}, $[N/2]$ denotes the integer part of $N/2$. 
\par Our expansion \eqref{approximation_weight_intro} and the related formulas are motivated by their two-dimensional prototypes of \cite{kunyansky1992generalized}, \cite{novikov2014weighted}, \cite{guillement2014inversion}.\\

\par In the present article we obtained,
in particular, the following results under assumptions \eqref{weight_func_assymp}, \eqref{w0_def}, \eqref{test_func_assymp}:
\begin{enumerate}
	\item If $\sigma_{\tW,D,\infty} < 1$, then $R_W$ is injective and, in addition, the inversion of $R_W$ is given via formulas \eqref{QWDinf_neym_series}, \eqref{inverse_rad_general_weighted}; see Subsection~\ref{subsect_precise_sol} for details.
	\item If $\sigma_{\tW,D,\infty} \geq 1$, then $f$ can be approximately reconstructed from $R_Wf$ as $f\approx (R_{\widetilde{W}_N})^{-1}R_Wf$, where $R_{\widetilde{W}_N}$ is defined according to \eqref{weighted_radon_general_def} for $\widetilde{W}_N$ defined 
	by \eqref{wN_tilde_def}
	for $N = 2m$, where $m$ is chosen as the largest while condition $\sigma_{\tW,D,m} < 1$ holds. More precisely, approximate inversion of $R_{W}f$ is given via the formulas \eqref{QWDm_neym_series}, \eqref{approx_m_def}; 
	see Subsection~\ref{subsect_diverg_case} for details.
	\par In addition, if $W = W_N$ defined by \eqref{wN_def} and $\sigma_{\tW,D,m} < 1, \, m = [N/2]$, then 
	$R_{W_N}$ is injective and invertible by formula \eqref{inverse_radon_weighted};
	see Subsection~\ref{subsect_finite_weight} for details.
\end{enumerate}
\par In addition, in these results assumptions \eqref{weight_func_assymp}, \eqref{w0_def}
can be relaxed as follows:
\begin{align}\label{new_weight_cond}
	&W\in L^{\infty}(\R^3\times \Sp^2), \\ \label{new_w0_cond}
	&w_{0,0} \geq c > 0 \text{ on } \R^3, 
\end{align}
where $w_{0,0}$ is defined \eqref{w0_def}, $c$ is some positive constant.
\par Prototypes of these results for the weighted Radon transforms in 2D were obtained in \cite{kunyansky1992generalized}, \cite{novikov2014weighted}, \cite{guillement2014inversion}.

\par The present work also continues studies of \cite{goncharov2016analog}, where approximate inversion of $R_W$ was realized as $(R_{W_N})^{-1}$ for $N = 0$ or by other words as an approximate Chang-type inversion formula. We recall that the original two-dimensional Chang formula (\cite{chang1978method}) is often used as an efficient first approximation in the framework of slice-by-slice reconstructions in the single photon emission computed tomography.

\par In Section~\ref{sect_prelim} we give some notations and preliminary results. 
\par The main results of the present work are presented in detail in Section~\ref{sect_main_results}.
\par In Section~\ref{sect_gen_multidim} we generalize results of Sections~\ref{sect_prelim},~\ref{sect_main_results} for the case of dimension $n > 3$.
\par Proofs of results of Sections~\ref{sect_prelim},
~\ref{sect_main_results}, ~\ref{sect_gen_multidim} are presented in Sections~\ref{sect_lemms_proofs},~\ref{sect_thm_proofs}.

\section{Some preliminary results}\label{sect_prelim}
\subsection{Some formulas for $R$ and $R^{-1}$}\label{subsect_prelim_radinv}
We recall that for the classical Radon transform $R$ (formula \eqref{weighted_radon_general_def} for $W\equiv 1$) the following identity holds (see \cite{natterer1986mathematics}, Theorem 1.2, p.13):
\begin{equation}\label{radon_conv_prop}
	R(f\ast_{\R^3}g) = Rf \ast_{\R} Rg,
\end{equation}
where $\ast_{\R^3}, \ast_{\R}$ denote the 3D and 1D convolutions (respectively), $f,g$ are test functions.
\par The classical Radon inversion formula is defined as follows (see, e.g.,  \cite{natterer1986mathematics}):
\begin{align}
	\label{radon_inv_3D}
	&R^{-1}q(x) = -\dfrac{1}{8\pi^2}
	\int\limits_{\Sp^{2}} q^{(2)}(x\theta, \theta)d\theta, \,
	 x\in \R^3,\\ \nonumber
	&q^{(2)}(s,\theta) = \dfrac{d^{2}}{ds^{2}} q(s,\theta), \, (s,\theta)\in \R\times \Sp^2, 
\end{align}
where $q$ is a test function on $\R\times \Sp^{2}$.
\par In addition, from the Projection theorem (see \cite{natterer1986mathematics}, Theorem 1.1, p.11) it follows that:
\begin{align}
	\label{rad_inverse_fourier_def}
	&R^{-1}q(x) \eqdef \dfrac{1}{(2\pi)^{5/2}}\int\limits_{\R}\dfrac{\rho^2}{2}d\rho\int\limits_{\Sp^2}
	\hat{q}(\rho, \omega)e^{i\rho(x\omega)}d\omega, \, x\in \R^3,\\
	&\hat{q}(s,\theta) \eqdef \dfrac{1}{\sqrt{2\pi}}\int\limits_{\R}q(t,\theta)e^{-its}\, dt, \,
	 \label{fourier_rad}
	  (s,\theta)\in \R\times\Sp^2, 
\end{align}
where $q(t,\theta)$ is a test function on $\R\times \Sp^2$.
\par For the case of $\hat{q}$ even (i.e., $\hat{q}(s,\theta) = \hat{q}(-s,-\theta), \, (s,\theta)\in \R\times \Sp^2$, where $\hat{q}$ is defined in \eqref{fourier_rad}), formulas \eqref{rad_inverse_fourier_def}, \eqref{fourier_rad} can be rewritten as follows:
\begin{align}\label{radon_fourier_conn}
	R^{-1}q = \dfrac{1}{2\pi}\mathcal{F}[\hat{q}] = \dfrac{1}{2\pi}\mathcal{F}^{-1}[\hat{q}],
\end{align}
where $\mathcal{F}[\cdot], \mathcal{F}^{-1}[\cdot]$ denote the Fourier transform and its inverse in 3D, respectively, and they are defined by the following formulas (in spherical coordinates):
\begin{align}\label{fourier_3d_straight}
	\mathcal{F}[q](\xi) &\eqdef \dfrac{1}{(2\pi)^{3/2}}\int\limits_{0}^{+\infty}\rho^2 d\rho
	\int\limits_{\Sp^2}q(\rho, \omega)e^{-i\rho(\xi\omega)}d\omega,\\
	\label{fourier_3d_inverse}
	\mathcal{F}^{-1}[q](\xi) &\eqdef \dfrac{1}{(2\pi)^{3/2}}\int\limits_{0}^{+\infty}\rho^2 d\rho
	\int\limits_{\Sp^2}q(\rho, \omega)e^{i\rho(\xi\omega)}d\omega, \, \xi \in \R^3,
\end{align}
where $q(\rho,\omega)$ is a test-function on $[0,+\infty)\times \Sp^2$ (identified with $\R^3$).

\subsection{Symmetrization of $W$}
Let
\begin{equation}\label{aw_def}
	A_Wf = R^{-1}R_Wf,
\end{equation}
where $R_W$ is defined in \eqref{weighted_radon_general_def}, $f$ is a test function, satisfying assumptions of \eqref{test_func_assymp}.
\par Let 
\begin{equation}\label{weight_symmetrization}
	\widetilde{W}(x,\theta) \eqdef \dfrac{1}{2}(W(x,\theta) + W(x,-\theta)), \, x\in \R^3, \theta \in \Sp^{2}.
\end{equation}
\par The following formulas hold:
\begin{align}
	\label{aw_symmetrization}
	&A_Wf = R^{-1}R_{\widetilde{W}}f,\\
	\label{radon_weight_symmetrization}
	&R_{\widetilde{W}}f(s,\theta)= \dfrac{1}{2}(R_Wf(s,\theta) + R_Wf(-s,-\theta)), \, (s,\theta)\in \R\times \Sp^{2},\\ 
	\label{weight_fourier_symmetrization}
	&\widetilde{W}(x,\theta(\gamma,\phi)) = \sum\limits_{k = 0}^\infty 
	\sum\limits_{n=-2k}^{2k}w_{2k,n}(x) Y_{2k}^n(\gamma,\phi),\\
	&x\in \R^3, \, \gamma\in [0,\pi], \, \phi\in[0,2\pi]. \nonumber
\end{align}
Identity \eqref{aw_symmetrization} is proved in \cite{goncharov2016analog} in 3D, where $\widetilde{W}$ is denoted as $W_{sym}$. 
\par Identity \eqref{weight_fourier_symmetrization} follows from \eqref{approximation_weight_intro}, \eqref{Ykn_def}, \eqref{weight_symmetrization}  and the following identities:
\begin{align}
	\label{spherical_symmetry_legendre}
	&p^{n}_k(-x) = (-1)^{n + k}p^{n}_k(x), \, x\in [-1,1],\\
	&e^{in(\phi + \pi)} = (-1)^n e^{in\phi}, \, k\in \mathbb{N}\cup \{0\}, \, 
	n = \overline{-k,k}.
	 \label{spherical_symmetry_exponent}
\end{align} 
\par Note also that $\widetilde{W}_N$ defined by \eqref{wN_tilde_def} is the approximation of $\widetilde{W}$ defined by \eqref{weight_symmetrization} and
\begin{equation}
	\widetilde{W}_N(x, \cdot ) \xrightarrow[N\rightarrow \infty]
	{L^2(\Sp^2)}
	\widetilde{W}(x, \cdot ) \text{ for each fixed }x\in \R^3. 
\end{equation}
\par Using formulas \eqref{radon_weight_symmetrization}-\eqref{weight_fourier_symmetrization} we reduce the inversion of $R_W$ to the inversion of $R_{\widetilde{W}}$ defined by \eqref{weighted_radon_general_def} for $W=\tW$. 
\par In our work $A_Wf$ (or, more precisely, $(w_{0,0})^{-1}A_Wf$)  is used as the \textit{initial point} for our iterative inversion algorithms (see Section~\ref{sect_main_results}). 
\par Note that the simmetrization $\tW$ of 
$W$ arises in \eqref{aw_symmetrization}.
\par In addition, prototypes of  \eqref{aw_def}, \eqref{aw_symmetrization}, \eqref{weight_fourier_symmetrization} for the two-dimensional case can be found in \cite{kunyansky1992generalized}, \cite{novikov2011weighted}.
\subsection{Operators $Q_{\tW,D,m}$ and numbers $\sigma_{\tW,D,m}$}\label{subsect_op_q}
Let
\begin{equation}
	\label{lower_bound_weight}
		c \eqdef \inf_{x\in D}|w_{0,0}(x)| > 0,
\end{equation}
where the inequality follows from the continuity of $W$ on $\overline{D}$ (closure of $D$) and assumption \eqref{w0_def}.\\

\noindent Let $D$ be the domain of \eqref{test_func_assymp}, and $\chi_D$ denote the characteristic function of $D$, i.e.
\begin{equation}
	\chi_D\equiv 1 \text{ on }D,\, \chi_D \equiv 0 \text{ on }\R^{3}\backslash D.
\end{equation}
Let 
\begin{align}\label{QWDm_def}
	&Q_{\tW,D,m}u(x) \eqdef R^{-1}(R_{\tW,D,m}u)(x), \, m\in \mathbb{N}\\ \nonumber
	&Q_{\tW,D,m}u(x) = 0 \text{ for }m = 0,\\ \label{QWDm_infty_def}
	&Q_{\tW,D,\infty}u(x) \eqdef R^{-1}(R_{\tW,D,\infty}u)(x), 
\end{align}
where
\begin{align}\label{RWDm_def}
	&R_{\tW,D,m}u(s,\theta(\gamma, \phi)) \eqdef \int\limits_{x\theta=s}\left(
		\sum\limits_{k=1}^{m}\sum\limits_{n=-2k}^{2k} \dfrac{w_{2k,n}(x)}{w_{0,0}(x)}
		Y_{2k}^{n}(\gamma,\phi)
	\right)\chi_D(x)u(x)\, dx,\\
	\label{RWDm_infty_def}
	&R_{\tW,D,\infty}u(s,\theta(\gamma, \phi)) \eqdef \lim\limits_{m\rightarrow \infty}
	R_{\tW,D,m}u(s,\theta(\gamma,\phi)) \nonumber\\
	&\qquad \qquad \qquad \qquad \quad = \int\limits_{x\theta=s}\left(
		\sum\limits_{k=1}^{\infty}
		\sum\limits_{n=-2k}^{2k} \dfrac{w_{2k,n}(x)}{w_{0,0}(x)}
		Y_{2k}^{n}(\gamma,\phi)
	\right)\chi_D(x)u(x)\, dx,\\ \nonumber
	&x\in \R^3,\, s\in \R, \, \theta(\gamma,\phi) \in \Sp^{2},
\end{align}
where $Y_{k}^{n}$ are defined in \eqref{Ykn_def}, $R^{-1}$ is defined by  \eqref{rad_inverse_fourier_def} (or \eqref{radon_inv_3D}), $u$ is a test function, $w_{0,0}, w_{2k,n}$ are the Fourier-Laplace coefficients defined by \eqref{wkn_def} and $w_{2k,n}/w_{0,0}, \chi_D$ are considered as multiplication operators on $\R^3$. 
Note also that $R_{\tW,D,\infty}(w_{0,0}f) + R(w_{0,0}f) = R_{\tW}f$ under assumptions \eqref{weight_func_assymp}-\eqref{test_func_assymp}.
\par Let 
\begin{equation}\label{d2kn_def}
	d_{2k,n}(x) \eqdef R^{-1}(\delta(\cdot)Y_{2k}^{n})(x), \, x\in \R^3, \, k\in \mathbb{N}, \, n = \overline{-2k,2k},
\end{equation}
where $\delta = \delta(s)$ denotes the 1D Dirac delta function. In \eqref{d2kn_def} the action of $R^{-1}$ on the generalized functions is defined by formula \eqref{rad_inverse_fourier_def}.

\begin{lemma}\label{lemma_exact_d2kn}
	Let $d_{2k,n}$ be defined by \eqref{d2kn_def}. Then 
	\begin{equation}\label{dkn_exact_def}
		d_{2k,n}(x(r,\gamma,\phi)) = \dfrac{(-1)^{k}\sqrt{2}\Gamma(\frac{3}{2} + k)}{\pi\Gamma(k)}
	\dfrac{Y_{2k}^n(\gamma,\phi)}{r^3}, r > 0,
	\end{equation}
	where $\Gamma(\cdot)$ is the Gamma-function, $x(r,\gamma,\phi)$ is defined by the identity:
	\begin{equation}\label{lemma_spherical_coord_def}
	x(r,\gamma, \phi) = (r\sin\gamma\cos\phi, r\sin\gamma\sin\phi, \, r\cos\gamma)\in \R^3, \, 
		\, \gamma \in [0,\pi], \, \phi \in [0,2\pi], \, r \geq 0.
	\end{equation}
	 In addition, the following inequality holds:
	\begin{equation}\label{lemma_fdkn_unifb}
		|\mathcal{F}[d_{2k,n}](\xi)| \leq \dfrac{1}{2\pi\sqrt{2}},\, \xi \in \R^3,
	\end{equation}
	where $\mathcal{F}[\cdot]$ is the Fourier transform, defined in \eqref{fourier_3d_straight}.
\end{lemma}
	\par The following lemma gives some useful expressions for operators $Q_{\tW, D,m}, \, Q_{\tW, D,\infty}$ defined in \eqref{QWDm_def}, \eqref{QWDm_infty_def}.

\begin{lemma}\label{lemma_qwd_useful}
	Let operators $Q_{\tW, D, m}, Q_{\tW, D,\infty}$ be defined by \eqref{QWDm_def}, \eqref{QWDm_infty_def},
	 respectively, and $u$ be a test function satisfying \eqref{test_func_assymp}. Then
	 \begin{align}\label{QWDm_conv_def}
	 	 &Q_{\tW,D,m}u = \sum\limits_{k=1}^{m}\sum\limits_{n = -2k}^{2k} 
	 	  d_{2k,n}\ast_{\R^3} \dfrac{w_{2k,n}}{w_{0,0}}
	 	  u,\\ \label{QWDinf_conv_def}
	 	 &Q_{\tW,D,\infty}u = \sum\limits_{k=1}^{\infty}\sum\limits_{n = -2k}^{2k} d_{2k,n}
	 	 \ast_{\R^3}\dfrac{w_{2k,n}}{w_{0,0}}u,
	 \end{align}
	where coefficients $w_{k,n}$ are defined in \eqref{wkn_def}, $d_{2k,n}$ is defined by
	\eqref{dkn_exact_def} (or equivalently by \eqref{d2kn_def}).
\end{lemma}
\begin{rem}
	Convolution terms in the right-hand side of \eqref{QWDm_conv_def}, \eqref{QWDinf_conv_def} are well defined functions in $L^2(\R^3)$. 
	This follows from identity \eqref{dkn_exact_def} and the Calder\'{o}n-Zygmund theorem for convolution-type operators with singular kernels (see \cite{knapp2005advanced}, p.83, Theorem 3.26).
\end{rem}

\par The following lemma shows that $Q_{\tW,D,m}, \, Q_{\tW,D,\infty}$ are well-defined operators in $L^2(\R^3)$.
\begin{lemma}\label{lemma_Qwdm_l2}
  Operator $Q_{\tW,D,m}$ defined by \eqref{QWDm_conv_def} (or equivalently by \eqref{QWDm_def}) is a linear bounded operator in $L^2(\R^3)$ and 
  the following estimate holds:
  \begin{equation}\label{QWDm_op_norm_bound}
		\|Q_{\tW,D,m}\|_{L^2(\R^3)\rightarrow L^2(\R^3)} \leq \sigma_{\tW,D,m},
	\end{equation}
	where $\sigma_{\tW,D,m}$ is defined by \eqref{sigma_WDm_def}.
	\par Operator $Q_{\tW,D,\infty}$ defined by \eqref{QWDinf_conv_def} (or equivalently by \eqref{QWDm_infty_def}), is a linear bounded operator in $L^2(\R^3)$ and the following estimate holds:
  \begin{equation}\label{QWDinf_op_norm_bound}
		\|Q_{\tW,D,\infty}\|_{L^2(\R^3)\rightarrow L^2(\R^3)} \leq \sigma_{\tW,D,\infty},
	\end{equation}
  where $\sigma_{\tW,D,\infty}$ is defined by \eqref{sigma_WDinfty_def}
\end{lemma}

\begin{lemma}\label{lemma2_RinvRW_glue}
Let 
\begin{equation}\label{lemma2_glue_assumption}
	\sum\limits_{k=1}^{\infty}\sum\limits_{n=-2k}^{2k}\left|\left|\dfrac{w_{2k,n}}{w_{0,0}}\right|\right|_{L^2(D)} < 
	+\infty,
\end{equation}
where $w_{k,n}$ are defined in \eqref{wkn_def}. Then 
\begin{equation}\label{lemma_l2_glue}
	R^{-1}R_Wf\in L^2(\R^3).
\end{equation}
In addition, the following formula holds:
\begin{equation}\label{lemma_radinv_conv}
	R^{-1}R_Wf = w_{0,0}f + \sum\limits_{k=1}^{\infty}\sum\limits_{n=-2k}^{2k} 
	d_{2k,n} \ast_{\R^3} w_{2k,n}f = (I + Q_{\tW,D,\infty})(w_{0,0}f),
\end{equation}
where $f$ satisfies \eqref{test_func_assymp}, operator $R^{-1}$ is defined by \eqref{rad_inverse_fourier_def} and $Q_{\tilde{W}, D,\infty}$ is given by \eqref{QWDinf_conv_def}.
\par In particular, if $W=W_N, \, N\in \mathbb{N}\cup \{0\}$, then the following analog of  \eqref{lemma_radinv_conv} holds:
\begin{equation}
	\label{lemma_radinv_conv_finite}
	R^{-1}R_Wf = w_{0,0}f + \sum\limits_{k=1}^{m}\sum\limits_{n=-2k}^{2k} 
	d_{2k,n} \ast_{\R^3} w_{2k,n}f = (I + Q_{\tW,D,m})(w_{0,0}f), \, m = [N/2],
\end{equation}
where $Q_{\tW, D, m}$ is given by \eqref{QWDm_conv_def}. 
\end{lemma}

\section{Main results}\label{sect_main_results}
\subsection{Case of $\sigma_{\tW, D,\infty} < 1$}\label{subsect_precise_sol}
Let 
\begin{equation}\label{sigma_WDinfty_conv}
	\sigma_{\tW,D,\infty} < 1,
\end{equation}
where $\sigma_{\tW,D,\infty}$ is defined by \eqref{sigma_WDinfty_def}. 
\par Inequality \eqref{sigma_WDinfty_conv} and estimate \eqref{QWDinf_op_norm_bound} in Lemma~\ref{lemma_Qwdm_l2} 
imply that operator $I + Q_{\tW,D,\infty}$ is continuosly invertible in $L^2(\R^3)$ and
the following identity holds (in the sense of the operator norm in $L^2(\R^3)$):
\begin{equation}\label{QWDinf_neym_series}
	(I + Q_{\tW,D,\infty})^{-1} = I + \sum\limits_{j=1}^{\infty} (-Q_{\tW,D,\infty})^j,
\end{equation}
where $I$ is the identity operator in $L^2(\R^3)$.
\begin{thm}\label{thm_infy_conv}
Let conditions \eqref{weight_func_assymp}-\eqref{test_func_assymp}, \eqref{sigma_WDinfty_conv} be fulfilled. Then $R_{W}$ defined by \eqref{weighted_radon_general_def} is injective and the following exact inversion formula holds:
\begin{equation}\label{inverse_rad_general_weighted}
	f = (w_{0,0})^{-1}(I + Q_{\tW,D,\infty})^{-1}R^{-1}R_Wf,
\end{equation}
where $w_{0,0}$ is defined in \eqref{w0_def}, $R^{-1}$ is defined in \eqref{rad_inverse_fourier_def}, 
operator $(I + Q_{\tW,D,\infty})^{-1}$ is given in \eqref{QWDinf_neym_series}.
\end{thm}
\begin{rem}
Formula \eqref{inverse_rad_general_weighted} can be considered as the following linear integral equation for the $w_{0,0}f$:
\begin{equation}\label{int_infWeight_equation}
	w_{0,0}f + Q_{\tW,D, \infty}(w_{0,0}f) = R^{-1}R_Wf.
\end{equation}
Inequality \eqref{sigma_WDinfty_conv} and identity \eqref{QWDinf_neym_series} imply that equation \eqref{int_infWeight_equation} can be solved by the method of successive approximations.
\end{rem}
\par One can see that, under conditions \eqref{weight_func_assymp}-\eqref{test_func_assymp} and \eqref{sigma_WDinfty_conv}, Theorem~\ref{thm_infy_conv} gives an exact inversion of $R_W$. However, condition \eqref{sigma_WDinfty_conv} is not always fulfilled in practice; see \cite{guillement2014inversion} for related numerical analysis in 2D. If condition \eqref{sigma_WDinfty_conv} is not fulfilled, then, approximating $W$ by finite Fourier series, in a similar way with \cite{chang1978method}, \cite{kunyansky1992generalized},  \cite{novikov2014weighted}, \cite{guillement2014inversion}, we suggest approximate inversion of $R_W$; see Subsections~\ref{subsect_diverg_case},~\ref{subsect_finite_weight}.

\subsection{Case of $1 \leq \sigma_{\tW,D,\infty} < +\infty$}\label{subsect_diverg_case}
Let
\begin{align}\label{sigma_WDappr_conv}
	&\sigma_{\tW,D,m} < 1, \text{ for some } m\in \mathbb{N}\cup \{0\},\\
	\label{full_weight_condition}
	&\sigma_{\tW, D,\infty} < +\infty,
\end{align}	
where $\sigma_{\tW,D,m}$ is defined by \eqref{sigma_WDm_def}, $\sigma_{\tW,D,\infty}$ is defined by \eqref{sigma_WDinfty_def}.
\par Inequality \eqref{sigma_WDappr_conv} and estimate \eqref{QWDm_op_norm_bound} in Lemma~\ref{lemma_Qwdm_l2} imply that $I+ Q_{\tW,D,m}$ is continuously invertible in $L^2(\R^3)$ and the following identity holds (in the sense of the operator norm in $L^2(\R^3)$):
\begin{equation}\label{QWDm_neym_series}
	(I + Q_{\tW,D,m})^{-1} = I + \sum\limits_{j=1}^{\infty}(-Q_{\tW,D,m})^j,
\end{equation}
where $I$ is the identity operator in $L^2(\R^3)$.
\begin{thm}\label{thm_infty_disconv}
 Let conditions \eqref{weight_func_assymp}-\eqref{test_func_assymp}, \eqref{sigma_WDappr_conv}, \eqref{full_weight_condition} be fulfilled. Then 
 	\begin{align}\label{approx_m_def}
 		&f \approx f_m \eqdef (w_{0,0})^{-1}(I + Q_{\tW,D,m})^{-1}R^{-1}R_Wf,\\
 		\label{approx_m_real_diff}
 		&f = f_m - (w_{0,0})^{-1}( I + Q_{\tW,D,m})^{-1}R^{-1}R_{\delta W_m}f,\\
 		\label{approx_m_norm_diff}
 		&\|f - f_m\|_{L^2(D)} \leq \dfrac{\|f\|_{\infty}}
 		{2\pi\sqrt{2} c(1 - \sigma_{\tW,D,m})} \sum\limits_{k = m + 1}^{\infty}
 		\sum\limits_{n = -2k}^{2k}\|w_{2k,n}\|_{L^2(D)} < + \infty,
 	\end{align}
 	where
	\begin{align}
	\label{delta_m_error}
		&\delta W_m(x,\theta(\gamma,\phi)) \eqdef W(x,\theta(\gamma,\phi)) - \sum\limits_{k = 0}^{2m+1}
		\sum\limits_{n = -k}^{k}w_{k,n}(x) Y^n_{k}(\gamma,\phi),\\
		&x\in \R^3, \, \gamma\in [0,\pi], \, \phi \in [0, 2\pi], \, 
		m\in \mathbb{N}\cup \{0\},
	\end{align}
	$w_{0,0}$ is defined in \eqref{w0_def}, 
	 $\theta(\gamma,\phi)$ is defined in \eqref{theta_gamma_phi},  
	 $Y_{k}^{n}$ are defined in \eqref{Ykn_def},
	 operator $(I + Q_{\tW,D,m})^{-1}$ is given in \eqref{QWDm_neym_series}, constant $c$ is defined in \eqref{lower_bound_weight}.
\end{thm}
\begin{rem}
	Formula \eqref{approx_m_def} can be considered as the following linear integral equation for $w_{0,0}f_m$:
	\begin{equation}\label{approx_m_equation}
		w_{0,0}f_m + Q_{\tW,D,m}(w_{0,0}f_m) = R^{-1}R_Wf.
	\end{equation}	 
	Inequality \eqref{sigma_WDappr_conv} and identity \eqref{QWDm_neym_series} imply that equation \eqref{approx_m_equation} is solvable by the method of successive approximations.
	\par Note also that condition \eqref{full_weight_condition} can be relaxed to
	the following one:
	\begin{equation}\label{cond_replacement}
		\sum\limits_{k=1}^{\infty}\sum\limits_{n=-2k}^{2k}\left|\left|\dfrac{w_{2k,n}}{w_{0,0}}\right|\right|_{L^2(D)} < 
	+\infty.
	\end{equation}
\end{rem}
Formula \eqref{approx_m_def} is an extension to 3D of the Chang-type two-dimensional inversion formulas in \cite{chang1978method}, \cite{novikov2014weighted},   \cite{guillement2014inversion}. In addition, formula \eqref{approx_m_def} is an extension of the approximate inversion formula in \cite{goncharov2016analog}, where this  formula was given for $m = 0$.
\par If \eqref{sigma_WDappr_conv} is fulfilled for some $m\geq 1$, then $f_m$ is a refinement of the Chang-type approximation $f_0$ and, more generally, $f_j$ is a refinement of $f_i$ for $0\leq i < j\leq m$. In addition, $f_j = f_i$ if $w_{2k,n}\equiv 0$ for $i < k\leq j, \, n = \overline{-2k, 2k}$. Thus, we propose the following approximate reconstruction of $f$ from $R_Wf$:
\begin{align*}
	&(i) \text{ \, find maximal }m\text{ such that \eqref{sigma_WDappr_conv} is still efficiently fulfilled},\\
	&(ii)\text{ approximately reconstruct }f \text{ by } f_m\text{ using \eqref{approx_m_def}}.
\end{align*}

\subsection{Exact inversion for finite Fourier series weights}\label{subsect_finite_weight}
\par Let 
\begin{equation}\label{weight_finite}
	W=W_N, \, N\in\mathbb{N}\cup \{0\},
\end{equation} 
where $W_N$ is defined by \eqref{wN_def}. 
\par Suppose that
\begin{equation}\label{QWDm_convergence}
	\sigma_{\tW,D,m} < 1\text{ for }m=[N/2],
\end{equation}
where $\sigma_{\tW,D,m}$ is defined by \eqref{sigma_WDm_def}.
\begin{thm}\label{thm_finite_appr}
Let conditions \eqref{weight_func_assymp}-\eqref{test_func_assymp}, \eqref{weight_finite}, \eqref{QWDm_convergence} be fulfilled. Then $R_{W}$ defined by \eqref{weighted_radon_general_def} is injective and the following exact inversion formula holds:
\begin{equation}\label{inverse_radon_weighted}
	f = (w_{0,0})^{-1}(I + Q_{\tW,D,m})^{-1}R^{-1}R_Wf,
\end{equation}
where $w_{0,0}$ is defined in \eqref{weight_func_assymp}, $(I + Q_{\tW,D,m})^{-1}$ is given in \eqref{QWDm_neym_series}, $R^{-1}$ is defined by \eqref{rad_inverse_fourier_def}.
\end{thm}
\begin{rem}
Formula \eqref{inverse_radon_weighted} can be considered as the following linear integral equation for $w_{0,0}f$:
\begin{equation}\label{int_finiteWeight_equation}
	w_{0,0}f + Q_{\tW,D,m}(w_{0,0}f) = R^{-1}R_Wf.
\end{equation}
 Identity \eqref{QWDm_convergence} implies that \eqref{int_finiteWeight_equation} can be solved by the method of successive approximations.
\end{rem}

\begin{rem}
	Note that Theorems~\ref{thm_infy_conv},~\ref{thm_infty_disconv},~\ref{thm_finite_appr} remain valid under
	assumptions \eqref{new_weight_cond}, \eqref{new_w0_cond} in place of \eqref{weight_func_assymp}, \eqref{w0_def}.
	This follows from the fact that in the proofs in Sections~\ref{sect_lemms_proofs},~\ref{sect_thm_proofs} it is required only existence of integral transforms, given by operators $R_W,\, R^{-1},\, \mathcal{F}[\cdot], \, \mathcal{F}^{-1}[\cdot], \,Q_{\tW,D,\infty}, \, Q_{\tW,D,m}$, their compositions (see Subsections~\ref{subsect_prelim_radinv},~\ref{subsect_op_q}) and of uniform upper bound on $(w_{0,0})^{-1}$ on $D$.
\end{rem}

\subsection{Additional comments}\label{subsect_add_comments}
\par There are different classes of weights $W$ for which the results of Subsections~\ref{subsect_precise_sol},~\ref{subsect_diverg_case},~\ref{subsect_finite_weight} can be used.\\
For example, condition \eqref{sigma_WDinfty_conv} is satisfied for the weights $W$ of the following form:
\begin{equation}\label{W_example}
	W(x,\theta) = c + V(x,\theta), \, (x,\theta)\in \R^3\times \Sp^2,
\end{equation}
where 
\begin{align}
	\begin{split}
	&c \text{ - is some positive constant}, \\	
	&V\in C^3(\R^3\times \Sp^2), \, \|V\|_{C^3(D\times\Sp^2)} \leq M(c,D),
	\end{split}
\end{align}
where $M$ is some positive constant depending only on $c$ and on $D$.
\par On the other hand, all weights $W$ admitting the finite Fourier series expansions
(i.e., $W=W_N$ of \eqref{wN_def} for some $N \in \mathbb{N}\cup \{0\}$) are dense in the space 
$C(\overline{D}\times \Sp^2)$ (and also in $L^2(D\times \Sp^2)$), where $\overline{D} = D\cup \partial D$. In addition, for such ``finite" $W_N\in C(\R^3\times \Sp^2)$ the sense of 
each of conditions \eqref{sigma_WDinfty_conv}, \eqref{sigma_WDappr_conv}, \eqref{full_weight_condition},
\eqref{cond_replacement}, \eqref{QWDm_convergence} is especially clear. Moreover, condition 
\eqref{full_weight_condition} is always satisfied, for example, if $W_N \in C(\R^3\times \Sp^2)$. In addition, even if \eqref{sigma_WDappr_conv} is not satisfied for the whole $W_N$, one can consider such a cutoff $W_m$ of $W_N, \, m < N$,  so that \eqref{sigma_WDappr_conv} holds and, therefore, 
results of Subsection~\ref{subsect_diverg_case} can be applied. 
\par Finally, we recall that in many cases even the zero order approximation $W \approx W_0 = w_{0,0}$ can be 
practically efficient in view of results presented in \cite{chang1978method}, \cite{novikov2011weighted}, 
\cite{guillement2014inversion}, \cite{goncharov2016analog}. Therefore, we expect that inversion algorithms 
of Subsections~\ref{subsect_diverg_case},~\ref{subsect_finite_weight} based on higher order finite Fourier approximations of $W\approx W_N$ may be even considerably more efficient in related tomographies.

\section{Generalization to multidimensions}\label{sect_gen_multidim}
\par Definition \eqref{weighted_radon_general_def} and assumptions \eqref{weight_func_assymp}-\eqref{test_func_assymp} are naturally extended as follows to the case of dimension $n>3$:
\begin{align}\label{weighted_radon_general_def_multidim}
	&R_Wf(s,\theta) = \int\limits_{x\theta = s} W(x,\theta) f(x)\, dx, \, (s,\theta) \in \R\times \Sp^{n-1}, \, x\in \R^n,\\ \label{weight_func_assymp_multidim}
	&W\in L^{\infty}(\R^n \times \Sp^{n-1}),\\ \label{w0_def_multidim}
	& w_{0,0}(x) \eqdef \dfrac{1}{|\Sp^{n-1}|}\int\limits_{\Sp^{n-1}} W(x,\theta) \, d\theta, \, w_{0,0} \geq c > 0 \text{ on } D, \\ \label{test_func_assymp_multidim}
	&f\in L^{\infty}(\R^n), \, \mathrm{supp}\, f \subset D,
\end{align}
where $|\Sp^{n-1}|$ denotes the standard Euclidean volume of $\Sp^{n-1}$, $c$ is a constant, 
$D$ is an open bounded domain in $\R^n$.
\par  For the weight $W$ we consider the Fourier-Laplace expansion:
\begin{equation}\label{approximation_weight_intro_multidim}
	W(x,\theta) = \sum\limits_{k = 0}^\infty 
	\sum\limits_{i=0}^{a_{k,n}-1}w_{k,i}(x) Y_k^i(\theta), \, x\in \R^n, \, \theta\in \Sp^{n-1},
\end{equation}
where
\begin{align} \label{fourier_coeff_multidim}
	&w_{k,i}(x) = \|Y_k^i\|_{L^2(\Sp^{n-1})}^{-2}\int\limits_{\Sp^{n-1}} W(x,\theta) \overline{Y_{k}^{i}(\theta)}\, d\theta, \,  \\
	\label{inner_dimension_def}
	&a_{k,n+1} = \dfrac{(n+k)!}{k!n!} - \dfrac{(n+k-2)!}{(k-2)!n!}, \, n,k \geq 2; \, a_{0,n}=1, \, a_{1,n}=n,
\end{align}
where $\{Y_k^i\, | \, k = \overline{0,\infty}, i=\overline{0,a_{k,n}-1}\}$ is the Fourier-Laplace basis of harmonics
on $\Sp^{n-1}$, $\overline{Y_{k}^{i}}$ denotes the complex conjugate of $Y_k^{i}$
; see \cite{stein2016introduction}, \cite{morimoto1998analytic}. In the present work we choose the basis $\{Y_{k}^{i}\}$ as in \cite{higuchi1987symmetric} without normalizing constants ${}_{n}c_{L}^{l}$ (i.e., $\{Y_k^i\}$ are the products of the Schmidt semi-normalized Legendre polynomials with one complex exponent and without any additional constants).\\

\noindent In dimension $n>3$, formulas \eqref{sigma_WDm_def}, \eqref{sigma_WDinfty_def}, \eqref{QWDm_def}, \eqref{QWDm_infty_def}, are rewritten as follows: 
\begin{align}\label{sigma_WDm_multidim_def}
	&\sigma_{\tW,D,m} \eqdef (2\pi\sqrt{2})^{(1-n)/2}\sum\limits_{k=1}^m \sum\limits_{i=0}^{a_{2k,n}-1}
	\sup\limits_{x\in D}\left|
		\dfrac{w_{2k,i}(x)}{w_{0,0}(x)}
	\right|,\\ \label{sigma_WDinf_multidim_def}
	&\sigma_{\tW,D,\infty} \eqdef \lim_{m\rightarrow +\infty}\sigma_{\tW,D,m}\\
	\label{QWDm_def_multidim}
	&Q_{\tW,D,m}u(x) \eqdef R^{-1}(R_{\tW,D,m}u)(x), \, m\in \mathbb{N}\\ \nonumber
	&Q_{\tW,D,m}u(x) = 0 \text{ for }m = 0,\\ \label{QWDinf_def_multidim}
	&Q_{\tW,D,\infty}u(x) \eqdef R^{-1}(R_{\tW,D,\infty}u)(x),
\end{align}
where
\begin{align}
	\label{RWDm_def_multidim}
	&R_{\tW,D,m}u(s,\theta) \eqdef \int\limits_{x\theta=s}\left(
		\sum\limits_{k=1}^{m}\sum\limits_{i=0}^{a_{2k,n}-1} \dfrac{w_{2k,i}(x)}{w_{0,0}(x)}Y_{2k}^{i}(\theta)
	\right)\chi_D(x)u(x)\, dx,\\ \label{RWDinfty_multidim_def}
	&R_{\tW,D,\infty}u(s,\theta) \eqdef \lim\limits_{m\rightarrow\infty} R_{\tW,D,m}u(s,\theta),\\
	\nonumber
	&x\in \R^n,\, s\in \R, \, \theta \in \Sp^{n-1},
\end{align}
where $R^{-1}$ is defined further in \eqref{radon_inverse_multidim}.

\par Under assumptions \eqref{weight_func_assymp_multidim}, \eqref{test_func_assymp_multidim}, for each fixed $x$, series of \eqref{approximation_weight_intro_multidim} converges in $L^2(\Sp^{n-1})$; see, e.g., \cite{stein2016introduction} (Chapter 4), \cite{morimoto1998analytic} (Chapter 2), \cite{zhizhiashvili1979fourier}.
\par Formula \eqref{rad_inverse_fourier_def} is extended as follows:
\begin{equation}\label{radon_inverse_multidim}
	R^{-1}q(x) = (2\pi)^{1/2-n}\int\limits_{\R}\dfrac{|\rho|^{n-1}}{2}\, d\rho
	\int\limits_{\Sp^{n-1}} \hat{q}(\rho,\theta)e^{i\rho(x\theta)} d\theta, \, x\in \R^n,
\end{equation}
where $q(s,\theta)$ is a test function on $\R\times\Sp^{n-1}$, $\hat{q}(s,\theta)$ is defined as in \eqref{fourier_rad} (with $\Sp^{n-1}$ in place of $\Sp^2$).
\par The Fourier transforms, defined in \eqref{fourier_3d_straight}, \eqref{fourier_3d_inverse}, are extended as follows:
\begin{align}\label{fourier_Rn}
	\mathcal{F}[q](\xi) &\eqdef (2\pi)^{-n/2}
	\int\limits_{0}^{+\infty}\rho^{n-1}d\rho\int\limits_{\Sp^{n-1}}
	q(\rho, \omega)e^{-i\rho (\xi\omega)}d\omega,\\
	\mathcal{F}^{-1}[q](\xi) &\eqdef (2\pi)^{-n/2} \label{fourier_Rn_inverse}
	\int\limits_{0}^{+\infty}\rho^{n-1}d\rho\int\limits_{\Sp^{n-1}}
	q(\rho, \omega)e^{i\rho (\xi\omega)}d\omega, \, \xi\in \R^n,
\end{align}
where $q(\rho,\omega)$ is a test function on $[0,+\infty)\times \Sp^{n-1}$ (identified with $\R^n$).

\par In dimension $n > 3$, formulas \eqref{aw_def}-\eqref{weight_fourier_symmetrization} remain valid with $Y_{k}^{m}$ defined in \eqref{Ykn_def} replaced by basis of spherical harmonics $\{Y_k^i\}$ on $\Sp^{n-1}$. In particular, the following multidimensional analog of \eqref{weight_fourier_symmetrization} holds:
\begin{equation}\label{harmonic_symmetry}
	Y_{k}^i(-\theta) = (-1)^kY_{k}^i(\theta), \, \theta\in \Sp^{n-1},
	\, k\in \mathbb{N}\cup \{0\}, \, i = \overline{0,a_{k,n}-1},
\end{equation}
where $a_{k,n}$ is defined by \eqref{inner_dimension_def}. Identity \eqref{harmonic_symmetry} reflects the fact that $Y_{k}^i(\theta)= Y_k^i(\theta_1,\theta_2, \cdots, \theta_n), \, \theta = (\theta_1, \theta_2, \cdots, \theta_n)\in \Sp^{n-1}, \, i=\overline{0,a_{k,n}-1}$ is a homogenous polynomial of degree $k$, see, e.g., \cite{stein2016introduction}, \cite{morimoto1998analytic}.\\

\noindent Formula \eqref{d2kn_def} is now rewritten as follows:
\begin{equation}\label{d2kn_def_multidim}
	d_{2k,i}(x) \eqdef R^{-1}(\delta(\cdot) Y_{2k}^{i})(x), \, x\in \R^n, \, i = \overline{0,a_{k,n}-1}.
\end{equation}
\noindent Results of Lemma~\ref{lemma_exact_d2kn} remain valid with formula \eqref{dkn_exact_def} replaced by the following one:
\begin{equation}\label{d2kn_exact_def_multidim}
	d_{2k,i}(r,\theta) = c(k,n) \dfrac{(-1)^k Y_{2k}^i(\theta)}{r^n}, \, r > 0, \, \theta\in \Sp^{n-1},
\end{equation}
where 
\begin{equation}\label{ckn_def}
	c(k,n) = \dfrac{\sqrt{2}\pi^{(1-n)/2} \Gamma(k + \frac{1}{2})
	\Gamma(k +\frac{n}{2})
	}
	{\Gamma(k)\Gamma(k + \frac{n-1}{2})}
	\cdot 
	\left(
	\dfrac{\Gamma(k + 1)}{\Gamma(k+\frac{1}{2})}
	\right)^{n-2},
\end{equation}
$\Gamma(\cdot)$ is the Gamma function.
\par In addition, inequality \eqref{lemma_fdkn_unifb} is rewritten as follows:
\begin{equation}
	|\mathcal{F}[d_{2k,i}](\xi)| \leq (2\pi\sqrt{2})^{(1-n)/2}, \, \xi \in \R^n,
\end{equation}
where $\mathcal{F}[\cdot]$ is the Fourier transform defined in \eqref{fourier_Rn}.
The constant $c(k,n)$ in \eqref{ckn_def} is obtained using formulas \eqref{radon_inverse_multidim}, \eqref{d2kn_def_multidim} and Theorems 1, 2 of \cite{goncharov2016spher}.
\par The results of Lemma~\ref{lemma_qwd_useful} remain valid in the case of dimension $n>3$, with  \eqref{QWDm_conv_def}, \eqref{QWDinf_conv_def} rewritten as follows:
\begin{align}
	\label{QWDm_conv_def_mdim}
	 &Q_{\tW,D,m}u = \sum\limits_{k=1}^{m}\sum\limits_{i = 0}^{a_{k,n}-1} 
	 d_{2k,i}\ast_{\R^n} \dfrac{w_{2k,i}}{w_{0,0}}
	 u,\\ \label{QWDinf_conv_def_mdim}
	 &Q_{\tW,D,\infty}u = \sum\limits_{k=1}^{\infty}\sum\limits_{i = 0}^{a_{k,n}-1} d_{2k,i}
	 \ast_{\R^n}\dfrac{w_{2k,i}}{w_{0,0}}u,
\end{align}
where coefficients $w_{2k,i},\,  w_{0,0}$ are defined in \eqref{fourier_coeff_multidim}, $a_{k,n}$ is defined in \eqref{inner_dimension_def}, $d_{2k,i}$ is defined in \eqref{d2kn_exact_def_multidim}, $\ast_{\R^n}$ denotes the convolution in $\R^n$.
\par The results of Lemma~\ref{lemma_Qwdm_l2} remain valid with $\R^3$ replaced by $\R^n,\, n>3$, where we use definitions \eqref{sigma_WDm_multidim_def}, \eqref{QWDm_def_multidim}, \eqref{QWDinf_def_multidim}.
\par Assumption \eqref{lemma2_glue_assumption} in Lemma~\ref{lemma2_RinvRW_glue} is rewritten now as follows:
\begin{equation}\label{lemm_glue_assump_multidim}
	\sum\limits_{k=1}^{\infty}
	\sum\limits_{i=0}^{a_{k,n}-1}
	\left|\left|
		\dfrac{w_{2k,i}}{w_{0,0}}
	\right|\right|_{L^2(D)} < +\infty.
\end{equation}
\par Under assumption \eqref{lemm_glue_assump_multidim}, property \eqref{lemma_l2_glue} of Lemma~\ref{lemma2_RinvRW_glue} remains valid in dimension $n>3$. In particular, formula \eqref{lemma_radinv_conv} is rewritten now as follows:
\begin{equation}\label{lemma_radinv_conv_multidim}
	R^{-1}R_Wf = w_{0,0}f + \sum\limits_{k=1}^{\infty}\sum\limits_{i=0}^{a_{k,n}-1} 
	d_{2k,i} \ast_{\R^n} w_{2k,i}f,
\end{equation}
where $R^{-1}$ is defined in \eqref{radon_inverse_multidim}, $f$ is a test function satisfying \eqref{test_func_assymp_multidim}.

\par Using formulas and notations from \eqref{approximation_weight_intro_multidim}-\eqref{QWDinf_conv_def_mdim} we obtain 
straightforward extensions of  Theorems~\ref{thm_infy_conv},~\ref{thm_infty_disconv},~\ref{thm_finite_appr}.

\begin{itemize}
\item[•]
The result of Theorem~\ref{thm_infy_conv} remains valid in dimension $n>3$, under assumptions \eqref{weight_func_assymp_multidim}-\eqref{test_func_assymp_multidim} and under condition \eqref{sigma_WDinfty_conv}, where 
 $w_{0,0}$ is defined in \eqref{w0_def_multidim}, 
 $R^{-1}$ is  defined in \eqref{radon_inverse_multidim}, 
 $\sigma_{\tW,D,\infty}$ is defined in \eqref{sigma_WDinf_multidim_def},
 $Q_{\tW,D,\infty}$ is defined in \eqref{QWDinf_def_multidim}.
 
\item[•] 

The result of Theorem~\ref{thm_infty_disconv} remains valid in dimension $n>3$,
under assumptions \eqref{weight_func_assymp_multidim}-\eqref{test_func_assymp_multidim} and under conditions \eqref{sigma_WDappr_conv}, \eqref{full_weight_condition},
where 
 $w_{0,0}$ is defined in \eqref{w0_def_multidim}, 
 $R^{-1}$ is  defined in \eqref{radon_inverse_multidim}, 
 $\sigma_{\tW,D,m}$ is defined in \eqref{sigma_WDm_multidim_def},
 $Q_{\tW,D,m}$ is defined in \eqref{QWDm_def_multidim} and where formulas 
 \eqref{approx_m_real_diff}-\eqref{delta_m_error} are rewritten as follows:
 \begin{align}
 		\label{approx_m_norm_diff_multidim}
 		&\|f - f_m\|_{L^2(D)} \leq \dfrac{\|f\|_{\infty}}
 		{(2\pi\sqrt{2})^{(n-1)/2} c(1 - \sigma_{\tW,D,m})} \sum\limits_{k = m + 1}^{\infty}
 		\sum\limits_{i = 0}^{a_{k,n}-1}\|w_{2k,i}\|_{L^2(D)} < + \infty,\\
		\label{delta_m_error_multidim}
		&\delta W_m(x,\theta) \eqdef W(x,\theta) - 	
		\sum\limits_{k = 0}^{2m+1}
		\sum\limits_{i = 0}^{a_{k,n}-1}w_{k,i}(x) Y^i_{k}(\theta),\\
		\nonumber
		&x\in \R^n, \, \theta\in \Sp^{n-1}.
 \end{align}

\item[•]
The result of Theorem~\ref{thm_finite_appr} remains valid in dimension $n>3$,
under assumptions \eqref{weight_func_assymp_multidim}-\eqref{test_func_assymp_multidim} and under conditions \eqref{weight_finite}, \eqref{QWDm_convergence},
where 
 $w_{0,0}$ is defined in \eqref{w0_def_multidim}, 
 $R^{-1}$ is  defined in \eqref{radon_inverse_multidim}, 
 $\sigma_{\tW,D,m}$ is defined in \eqref{sigma_WDm_multidim_def},
 $Q_{\tW,D,m}$ is defined in \eqref{QWDm_def_multidim}.
\end{itemize}
\vspace*{0.3cm}

\noindent The related proofs are the straightforward extensions to the case of dimension $n>3$ of proofs in Section~\ref{sect_thm_proofs} for $n=3$.

\section{Proofs of Lemma~\ref{lemma_exact_d2kn},~\ref{lemma_qwd_useful},~\ref{lemma_Qwdm_l2},~\ref{lemma2_RinvRW_glue}}\label{sect_lemms_proofs}
\subsection{Proof of Lemma~\ref{lemma_exact_d2kn}}

We consider $x(r,\gamma, \phi)$ defined by \eqref{lemma_spherical_coord_def} and $\omega(\gamma,\phi) = x(1,\gamma,\phi)$ (i.e., $\omega\in \Sp^2$).
\par Identity \eqref{lemma_spherical_coord_def} implies the following expression for the 
scalar product $(x\omega)$:
\begin{equation}\label{scal_prod}
	(x\omega) = (x(r,\tilde{\gamma},\tilde{\phi}), \omega(\gamma,\phi)) = 
	r(\cos\gamma\cos \tilde{\gamma} + \sin\gamma\sin\tilde{\gamma}\cos(\phi-\tilde{\phi})),
\end{equation}
where $\gamma, \tilde{\gamma} \in [0,\pi], \, \phi, \tilde{\phi}\in [0,2\pi],\, r\geq 0$.
\par From formulas \eqref{Ykn_def}, \eqref{rad_inverse_fourier_def}, \eqref{d2kn_def}, \eqref{scal_prod} it follows that
\begin{align}\nonumber
	(2\pi)^{5/2}d_{2k,n}(x(r,\tilde{\gamma},\tilde{\phi})) &= \int\limits_{\R}\dfrac{\rho^2}{2}d\rho \int\limits_{\Sp^2} 
	e^{i\rho(x\omega(\gamma,\phi))}
	Y^{n}_{2k}(\gamma,\phi) \, d\omega(\gamma,\phi)\\ \nonumber
	&= \int\limits_{\R}\dfrac{\rho^2}{2} d\rho \int\limits_{0}^{\pi}\sin(\gamma)
	 \, p_{2k}^{|n|}(\cos\gamma) \, d\gamma
	\int\limits_{0}^{2\pi}e^{i\rho(x\omega(\gamma,\phi)) + in\phi} d\phi \\ 
	\nonumber
	&= e^{in\tilde{\phi}}\int\limits_{\R}\dfrac{\rho^2}{2}d\rho
	\int\limits_{0}^{\pi} \sin(\gamma) p_{2k}^{|n|}(\cos\gamma) 
	e^{i\rho r \cos \gamma\cos \tilde{\gamma}}\,
	d\gamma
	\int\limits_{0}^{2\pi} 
	e^{i\rho r\sin\gamma \sin\tilde{\gamma}\cos(\phi-\tilde{\phi}) + 
	in(\phi - \tilde{\phi})} d\phi\\ \nonumber
	&= e^{in\tilde{\phi}}\int\limits_{\R}\dfrac{\rho^2}{2}d\rho
	\int\limits_{0}^{\pi} \sin(\gamma) p_{2k}^{|n|}(\cos\gamma) 
	e^{i\rho r \cos \gamma\cos \tilde{\gamma}}\, d\gamma
	\int\limits_{0}^{2\pi} 
	e^{i\rho r\sin\gamma \sin\tilde{\gamma}\cos\phi + in\phi} d\phi
	\\
    \label{integrand_gamma}
	&=2\pi e^{in(\tilde{\phi}-\pi/2)}(-1)^{n}\int\limits_{\R}\dfrac{\rho^2}{2}d\rho
	\int\limits_{0}^\pi
		 \sin(\gamma)p_{2k}^{|n|}(\cos\gamma)
	e^{i\rho r\cos\gamma \cos\tilde{\gamma}} 
	J_n(\rho r \sin\gamma\sin\tilde{\gamma})\, d\gamma,
\end{align}
where $J_n$ is the $n$-th standard Bessel function of the first kind; see e.g. \cite{temme2011special}. In \eqref{integrand_gamma} we used the well known formula for the Bessel function $J_n$: 
\begin{align}\nonumber
	J_n(t) \eqdef \dfrac{1}{2\pi}\int\limits_{-\pi}^{\pi}e^{in\phi - t\sin\phi}
	d\phi = \dfrac{(-1)^n e^{in\pi/2}}{2\pi} 
	\int\limits_{0}^{2\pi}e^{in\phi + t\cos\phi}d\phi.
\end{align}
\par The integral in $d\gamma$ in the right-hand side of \eqref{integrand_gamma}
was considered in \cite{neves2006analytical}, where the following exact analytic solution was given:
\begin{equation}\label{bessel_integration}
	\int\limits_{0}^\pi
		 \sin(\gamma)p_{2k}^{|n|}(\cos\gamma)
	e^{i\rho r\cos\gamma \cos\tilde{\gamma}} 
	J_n(\rho r \sin\gamma\sin\tilde{\gamma})\, d\gamma = 2i^{2k-n}p_{2k}^{|n|}(\cos \tilde{\gamma})j_{2k}(\rho r),
\end{equation}
where $j_{2k}$ is the standard spherical Bessel function of order $2k$; see e.g. \cite{temme2011special}.
\par From identities \eqref{integrand_gamma}, \eqref{bessel_integration}
it follows that:
\begin{align}\label{lemma_d2kn_exact_proof}\nonumber
	d_{2k,n}(x(r,\tilde{\gamma}, \tilde{\phi})) &= \dfrac{(-1)^{k}}{(2\pi)^{3/2}}
	p_{2k}^{|n|}(\cos\tilde{\gamma})e^{in\tilde{\phi}}
	\int\limits_{\R}\rho^2 j_{2k}(\rho r)\, d\rho\\
	&= \dfrac{4\sqrt{\pi}(-1)^{k}\Gamma(\frac{3}{2} + k)}{(2\pi)^{3/2}\Gamma(k)}
	\dfrac{p_{2k}^{|n|}(\cos\tilde{\gamma})e^{in\tilde{\phi}}}{r^3}, \, r > 0.
\end{align}
where $\Gamma(\cdot)$ is the Gamma function.
\par Definition \eqref{Ykn_def} and identity \eqref{lemma_d2kn_exact_proof} imply formula \eqref{dkn_exact_def}. \\
Formulas \eqref{radon_fourier_conn}, \eqref{spherical_symmetry_legendre}, \eqref{spherical_symmetry_exponent}, \eqref{d2kn_def} imply that 
\begin{align}\label{dkn_fourier_form}
(2\pi)d_{2k,n}(x(r,\tilde{\gamma},\tilde{\phi})) = 
 \mathcal{F}^{-1}[Y^{n}_{2k}](x(r, \tilde{\gamma}, \tilde{\phi})), 
	\, r > 0, \, \tilde{\gamma}\in [0,\pi], \, \tilde{\phi} \in [0,2\pi], 
\end{align} 
where $\mathcal{F}^{-1}[\cdot]$ is defined in \eqref{fourier_3d_inverse}.
\par Due to invertibility of the Fourier transform defined in \eqref{fourier_3d_straight} and identity \eqref{dkn_fourier_form} the following identity holds: 
\begin{equation}
2\pi\mathcal{F}[d_{2k,n}](\xi) = \mathcal{F}\mathcal{F}^{-1}[Y_{2k}^{n}](\xi/ |\xi|) = Y_{2k}^{n}(
\xi/|\xi|), \, \xi \in \R^3\backslash \{0\}.
\end{equation}
\par For $Y_{k}^n$ defined in \eqref{Ykn_def} the following inequality holds (see, e.g.,  \cite{lohofer1998legendre}):
\begin{equation}\label{legendre_exp_upbound}
	|Y_{k}^n(\gamma,\phi)| \leq \dfrac{1}{\sqrt{2}}, \, \gamma\in [0,\pi],\, \phi\in [0,2\pi].
\end{equation}
\par Identities \eqref{dkn_fourier_form} and inequality \eqref{legendre_exp_upbound} imply \eqref{lemma_fdkn_unifb}.
\par Note that $|\mathcal{F}[d_{2k,n}](\xi)|$, is uniformly bounded by $1/(2\pi\sqrt{2})$ everywhere except only at one point $\xi = 0$, where direction $\xi/|\xi|\in \Sp^2$ is not defined. However, point $\xi=0$ is of Lebesgue measure zero and $\mathcal{F}[d_{2k,n}]$ can be defined with any value at the origin in $\R^3$. 
\par Lemma~\ref{lemma_exact_d2kn} is proved.

\subsection{Proof of Lemma~\ref{lemma_qwd_useful}}
From identity \eqref{QWDm_def} it follows that 
\begin{align}\nonumber
	Q_{\tW,D,m}u &= R^{-1}\left(
		\sum\limits_{k=1}^{m}\sum\limits_{n = -2k}^{2k}
		Y_{2k}^{n} R\left(
			\dfrac{w_{2k,n}}{w_{0,0}}\chi_D u
		\right)
	\right)\\ \label{lemm1_proof_qwdm_ident}
	&= R^{-1}\left(
		\sum\limits_{k=1}^{m}\sum\limits_{n = -2k}^{2k}
		(\delta(\cdot)Y_{2k}^{n}) \ast_{\R}
		R\left(
			\dfrac{w_{2k,n}}{w_{0,0}}\chi_D u
		\right)
	\right)\\ \nonumber
	&= R^{-1}\left(
		\sum\limits_{k=1}^{m}\sum\limits_{n = -2k}^{2k}
		R(d_{2k,n}) \ast_{\R}
		R\left(
			\dfrac{w_{2k,n}}{w_{0,0}}\chi_D u
		\right)
	\right)
\end{align}
where $\ast_{\R}$ denotes the 1D convolution, $\delta = \delta(s)$ is the 1D Dirac delta function, $d_{2k,n}$ is defined by \eqref{d2kn_def}.
\par Identities \eqref{radon_conv_prop}, \eqref{lemm1_proof_qwdm_ident} imply \eqref{QWDm_conv_def}.
\par For the operator $Q_{\tW, D,\infty}$ defined by \eqref{QWDm_infty_def} we proceed 
according to identity \eqref{lemm1_proof_qwdm_ident} with $m\rightarrow +\infty$. Identites \eqref{radon_conv_prop}, \eqref{lemm1_proof_qwdm_ident} and linearity of operator $R^{-1}$ defined by \eqref{rad_inverse_fourier_def}  imply \eqref{QWDinf_conv_def}.
\par Lemma~\ref{lemma_qwd_useful} is proved.

\subsection{Proof of Lemma~\ref{lemma_Qwdm_l2}}
\par From formula \eqref{QWDm_conv_def} and the fact that the Fourier transform defined in \eqref{fourier_3d_straight} does not change the $L^2$-norm we obtain:
\begin{align}\label{lemm1_qwdm_four_norm} \nonumber
	\|Q_{\tW,D,m}u\|_{L^2(\R^3)} &\leq
	 \sum\limits_{k=1}^{m}\sum_{n=-2k}^{2k} 
	 \left|\left|
	 	d_{2k,n}\ast_{\R^3}\dfrac{w_{2k,n}}{w_{0,0}}\chi_Du
	 \right|\right|_{{L^2}(\R^3)}\\
	& = \sum\limits_{k=1}^{m}\sum_{n=-2k}^{2k}
	\left|\left|\mathcal{F}[d_{2k,n}]
	\mathcal{F}\left[\dfrac{w_{2k,n}}{w_{0,0}}\chi_Du 
	\right]
	\right|\right|_{{L^2}(\R^3)}.
\end{align}
\par From inequalities \eqref{lemma_fdkn_unifb}, \eqref{lemm1_qwdm_four_norm} we obtain:
\begin{align}\label{lemm1_qwdm_sigma_estim}
	\|Q_{\tW,D,m}u\|_{L^2(\R^3)} &\leq \dfrac{1}{2\pi\sqrt{2}} \nonumber
	\sum\limits_{k=1}^{m}\sum\limits_{n=-2k}^{2k}\left|\left|
		\mathcal{F}\left(\dfrac{w_{2k,n}}{w_{0,0}}\chi_D u 
	\right)
	\right|\right|_{L^2(\R^3)} \\ \nonumber
	&= \dfrac{1}{2\pi\sqrt{2}}
	\sum\limits_{k=1}^{m}\sum\limits_{n=-2k}^{2k}\left|\left|
		\dfrac{w_{2k,n}}{w_{0,0}}\chi_D u 
	\right|\right|_{L^2(\R^3)}\\
	&\leq \sigma_{\tW,D,m}\|u\|_{L^2(D)},
\end{align}
where $\sigma_{\tW,D,m}$ is defined by \eqref{sigma_WDm_def}. 
\par Inequality \eqref{lemm1_qwdm_sigma_estim} implies \eqref{QWDm_op_norm_bound}.
\par Estimate \eqref{QWDinf_op_norm_bound} follows from definition \eqref{QWDm_infty_def}, formula \eqref{QWDinf_conv_def}, linearity of operator $R^{-1}$ defined by \eqref{radon_inv_3D} and  inequalities 
\eqref{lemm1_qwdm_four_norm}, \eqref{lemm1_qwdm_sigma_estim} for $m\rightarrow +\infty$.
\par Lemma~\ref{lemma_Qwdm_l2} is proved.

\subsection{Proof of Lemma~\ref{lemma2_RinvRW_glue}}
\par From formulas \eqref{weighted_radon_general_def}, \eqref{approximation_weight_intro}
it follows that
\begin{align}\label{lemm2_Rw_def}
	&R_Wf(s,\theta(\gamma,\phi)) = \sum\limits_{k = 0}^{\infty}
	\sum\limits_{n = -k}^{k}
	Y_{k}^{n}(\gamma,\phi) R(w_{k,n}f)(s,\theta(\gamma,\phi)),\\
	&s\in \R, \, \gamma \in [0,\pi], \, \phi\in [0,2\pi],
\end{align}
where $\theta(\gamma, \phi)$ is defined in \eqref{theta_gamma_phi},
$Y_{k}^{n}(\gamma,\phi)$ are defined by \eqref{Ykn_def}, $w_{k,n}$ are defined in
 \eqref{wkn_def}.
\par Formula \eqref{lemma_radinv_conv} follows from formulas \eqref{aw_def}, \eqref{aw_symmetrization}, \eqref{weight_fourier_symmetrization}, \eqref{QWDm_infty_def} and formula \eqref{QWDinf_conv_def} in Lemma~\ref{lemma_qwd_useful}, where test function $u$ is replaced by $w_{0,0}u$.
\par From inequality \eqref{lemma_fdkn_unifb}, formulas \eqref{radon_conv_prop}, \eqref{lemma_radinv_conv} and the fact that the Fourier tranfsform defined in \eqref{fourier_3d_straight} does not change the $L^2$-norm we obtain:
\begin{align}\label{lemm2_RinvRw_norm}
	\|R^{-1}R_Wf\|_{L^2(\R^3)} &\leq \nonumber
	\|w_{0,0}f\|_{L^2(\R^3)} + \sum \limits_{k=1}^{\infty}\sum\limits_{n=-2k}^{2k}
	\|d_{2k,n}\ast_{\R^3}w_{2k,n}f\|_{L^2(\R^3)}\\
	&= \|w_{0,0}f\|_{L^2(\R^3)} + \sum \limits_{k=1}^{\infty}\sum\limits_{n=-2k}^{2k}
	\|\mathcal{F}[d_{2k,n}]\mathcal{F}[w_{2k,n}f]\|_{L^2(\R^3)}\\
	&\leq \|w_{0,0}f\|_{L^2(\R^3)} + \nonumber
	\dfrac{\|f\|_{\infty}}{2\pi\sqrt{2}}\sum \limits_{k=1}^{\infty}\sum\limits_{n=-2k}^{2k}
	\|w_{2k,n}\|_{L^2(D)},
\end{align}
where $\|\cdot \|_{\infty}$ denotes the $L^{\infty}$-norm.
\par From assumption \eqref{lemma2_glue_assumption} and formula \eqref{lower_bound_weight} it follows that
\begin{equation}\label{lemm2_nomAssump_conv}
	\sum\limits_{k=1}^{\infty}\sum\limits_{n=-2k}^{2k} \|w_{2k,n}\|_{L^2(D)} < + \infty.
\end{equation}
\par Assumptions  \eqref{weight_func_assymp}-\eqref{test_func_assymp} and inequalities \eqref{lemm2_RinvRw_norm}, \eqref{lemm2_nomAssump_conv} imply that
\begin{equation}\label{lemm2_final_bound}
	\|R^{-1}R_Wf\|_{L^2(\R^3)} \leq \|w_{0,0}f\|_{L^2(\R^3)} + 
	\dfrac{\|f\|_{\infty}}{2\pi\sqrt{2}}\sum\limits_{k=1}^{\infty}\sum\limits_{n=-2k}^{2k}
	\|w_{2k,n}\|_{L^2(D)} < + \infty.
\end{equation}
\par Lemma 2 is proved.

\section{Proofs of Theorems~\ref{thm_infy_conv},~\ref{thm_infty_disconv},~\ref{thm_finite_appr}}\label{sect_thm_proofs}

\subsection{Proof of Theorem~\ref{thm_infy_conv}}
\par Property \eqref{lemma_l2_glue} and identity \eqref{lemma_radinv_conv} of Lemma~\ref{lemma2_RinvRW_glue} follow from assumption \eqref{sigma_WDinfty_conv}. 
\par Formulas \eqref{lemma_l2_glue}, \eqref{lemma_radinv_conv}, \eqref{sigma_WDinfty_conv}, \eqref{QWDinf_neym_series} imply formula  \eqref{inverse_rad_general_weighted}.
\par The injectivity of $R_W$
follows from formula \eqref{inverse_rad_general_weighted}.
\par Theorem~\ref{thm_infy_conv} is proved.

\subsection{Proof of Theorem~\ref{thm_finite_appr}}
Property \eqref{lemma_l2_glue} and identity \eqref{lemma_radinv_conv_finite} of Lemma~\ref{lemma2_RinvRW_glue} follow from assumption \eqref{QWDm_convergence}. 
\par Formulas \eqref{lemma_l2_glue}, \eqref{lemma_radinv_conv_finite}, \eqref{QWDm_neym_series}, \eqref{QWDm_convergence} imply formula  \eqref{inverse_radon_weighted}.
\par The injectivity of $R_W$ follows from formula \eqref{inverse_radon_weighted}.
\par Theorem~\ref{thm_finite_appr} is proved.

\subsection{Proof of Theorem~\ref{thm_infty_disconv}}
Inequality \eqref{lemma2_glue_assumption} follows from assumption \eqref{full_weight_condition}. Hence, formulas \eqref{lemma_l2_glue}, \eqref{lemma_radinv_conv} (in Lemma~\ref{lemma2_RinvRW_glue}) hold.
\par Assumptions \eqref{weight_func_assymp}-\eqref{test_func_assymp},  \eqref{sigma_WDappr_conv} and inequality \eqref{QWDm_op_norm_bound} from Lemma~\ref{lemma_Qwdm_l2} imply that $f_m\in L^2(\R^3)$, where $f_m$ is defined in \eqref{approx_m_def}.
\par We split expansion \eqref{approximation_weight_intro} of weight $W$ defined by \eqref{weight_func_assymp} in the following way:
\begin{align}\label{thm_idcnv_weight_split}
	W(x,\theta) = W_{N + 1}(x,\theta) + \delta W_m(x,\theta), \, \theta\in \Sp^2, \, x\in \R^3, \,  m = [N/2], 
\end{align}
where $W_{N + 1}$ is defined by \eqref{wN_def}, $[N/2]$ denotes the integer part of $N/2$, $\delta W_m$ is defined by \eqref{delta_m_error}. 
\par From \eqref{weight_func_assymp}-\eqref{test_func_assymp} and from  \eqref{thm_idcnv_weight_split} it follows that 
\begin{equation}\label{thm_idcnv_radon_split}
	R_Wf = R_{W_{N+1}}f + R_{\delta W_m}f,
\end{equation}
where $R_Wf, R_{W_{N+1}}f, R_{\delta W_m}f$ are defined by \eqref{weighted_radon_general_def} for the case of weights $W, W_{N+1}, \delta W_m$ defined in \eqref{weight_func_assymp}, \eqref{wkn_def}, \eqref{delta_m_error}, respectively.
\par Identity \eqref{thm_idcnv_radon_split}
implies that 
\begin{equation}\label{thm_idcnv_inv_radon_split}
	R^{-1}R_Wf = R^{-1}R_{W_N}f + R^{-1}R_{\delta W_m}f,
\end{equation}
where $R^{-1}$ is defined by \eqref{rad_inverse_fourier_def}. Inequality \eqref{lemma2_glue_assumption} for the cases of weights $W, \, W_{N}, \, \delta W_m$, follows from assumption \eqref{full_weight_condition}. 
Therefore, Lemma~\ref{lemma2_RinvRW_glue} holds for $W, \, W_{N}, \, \delta W_m$ and, in particular, from \eqref{lemma_l2_glue} we have that:
\begin{equation}\label{thm_idcnv_inv_split_l2}
	R^{-1}R_{W_N}f\in L^2(\R^3), \, R^{-1}R_{\delta W_m}f\in L^2(\R^3).
\end{equation}
Theorem~\ref{thm_finite_appr} for $W=W_N, \, N = 2m$ holds by assumption \eqref{sigma_WDappr_conv}. Therefore, from formula \eqref{inverse_radon_weighted} we obtain:
\begin{equation}\label{thm_idcnv_fin_term}
	f = (w_{0,0})^{-1}(I + Q_{\tW,D,m})^{-1}R^{-1}R_{W_N}f,
\end{equation}
where operator $Q_{\tW,D,m}$ is defined in \eqref{QWDm_def} for $m$ arising in \eqref{sigma_WDappr_conv}.

From \eqref{approx_m_def}, \eqref{thm_idcnv_inv_radon_split}, \eqref{thm_idcnv_inv_split_l2}, \eqref{thm_idcnv_fin_term} it follows that:
\begin{align}
\begin{split}\label{ref_approx_mreal_diff}
	f_m &= (w_{0,0})^{-1}(I + Q_{\tW,D,m})^{-1}R^{-1}R_{W}f \\
	&= (w_{0,0})^{-1}(I + Q_{\tW,D,m})^{-1}R^{-1}R_{W_N}f\\
	&+ (w_{0,0})^{-1}(I + Q_{\tW,D,m})^{-1}R^{-1}R_{\delta W_m}f\\
	&= f + (w_{0,0})^{-1}(I + Q_{\tW,D,m})^{-1}R^{-1}R_{\delta W_m}f.
\end{split}
\end{align}
\par Formula \eqref{approx_m_real_diff} directly follows from \eqref{ref_approx_mreal_diff}. 

\par Inequality \eqref{sigma_WDappr_conv} and identities \eqref{QWDm_neym_series}, \eqref{approx_m_real_diff} imply the following inequality:
\begin{equation}\label{thm_idcnv_final_bound}
	\|f - f_m\|_{L^2(\R^3)} \leq \dfrac{1}{c}
	\|(I + Q_{\tW,D,m})^{-1}\|_{L^2(\R^3)\rightarrow L^2(\R^3)}\cdot 
	\|R^{-1}R_{\delta W_m}f\|_{L^2(\R^3)}, 
\end{equation}
where $c$ is defined in \eqref{lower_bound_weight}, $Q_{\tW,D,m}$ is defined by \eqref{QWDm_def} for $m$ in \eqref{sigma_WDappr_conv}.
\par From \eqref{QWDm_op_norm_bound} of Lemma~\ref{lemma_Qwdm_l2} and from identity \eqref{QWDm_neym_series} it follows that:
\begin{equation}\label{geometric_summ_norm}
	\|(I + Q_{\tW,D,m})^{-1}\|_{L^2(\R^3)\rightarrow L^2(\R^3)}
	\leq 1 + \sum\limits_{j=1}^{\infty} \|Q_{\tW,D,m}\|^j_{L^2(\R^3)\rightarrow L^2(\R^3)} \leq 
	\dfrac{1}{1-\sigma_{\tW,D,m}}.
\end{equation}
\par From formulas \eqref{lemma_radinv_conv}, \eqref{delta_m_error} and according to \eqref{lemm2_RinvRw_norm} it follows that 
\begin{equation}\label{RinvRdm_op_norm}
	\|R^{-1}R_{\delta W_m}f\|_{L^2(\R^3)} \leq \dfrac{\|f\|_{\infty}}{2\pi\sqrt{2} }\sum\limits_{k = m + 1}^{\infty}
	\sum\limits_{n = -2k}^{2k} \|w_{2k,n}\|_{L^2(D)},
\end{equation} 
where $R^{-1}$ is defined by \eqref{rad_inverse_fourier_def}, $w_{2k,n}$ are defined by \eqref{wkn_def}.
\par Putting the estimates \eqref{geometric_summ_norm}, \eqref{RinvRdm_op_norm} in the right-hand side of  \eqref{thm_idcnv_final_bound} we obtain \eqref{approx_m_norm_diff}.
\par Theorem~\ref{thm_infty_disconv} is proved. 

\section{Aknowledgments}
The present work was fulfilled in the framework of research conducted under the direction of Prof. R.G. Novikov. The author is also grateful to the referees for remarks that have helped to improve the presentation. This work is partially supported by the PRC $n^{\circ}$ 1545 CNRS/RFBR: \'{E}quations quasi-lin\'{e}aires, probl\`{e}mes inverses et leurs applications.


\end{document}